# Instructional Goals and Grading Practices of Graduate Students after One Semester of Teaching Experience

Charles Henderson[1], Emily Marshman[2], Alexandru Maries[2], Edit Yerushalmi[3], and Chandralekha Singh[2]

[1] *Department of Physics, Western Michigan University, 1903 W. Michigan Ave., Kalamazoo, MI, 49008*
[2] *Department of Physics and Astronomy, University of Pittsburgh, 3941 O'Hara St., Pittsburgh, PA 15260*
[3] *Department of Science Teaching, Weizmann Institute of Science, 234 Herzl St. Rehovot, Israel 7610001*

**Abstract:** Teaching assistants (TAs) are often responsible for grading student solutions. Since grading communicates instructors' expectations, TAs' grading decisions play a crucial role in forming students' approaches to problem solving (PS) in physics. We investigated the change in grading practices and considerations of 18 first-year graduate students participating in a TA professional development (PD) course. The TAs were asked to state their beliefs about the purpose of grading, to grade a set of specially designed student solutions, and to explain their grading decisions. We found that after one semester of teaching experience and participation in PD, TAs did not significantly change their goals for grading (i.e., a learning opportunity for both the student and the instructor) or their grading practice. In addition, TAs' grading practice frequently did not align with their goals. However, some TAs' perceptions of the level of explication required in a student solution did change. Our findings suggest that in order for PD to help TAs better coordinate their goals with appropriate grading practices, PD should focus on TAs' perception of sufficient reasoning in student solutions.

## INTRODUCTION

Problem solving (PS) plays a central role in physics teaching. Instructors require physics students to solve problems both to improve PS skills (i.e., developing expert-like approaches to PS) as well as to develop conceptual understanding of physics topics [1-4]. Research has shown that it is possible to advance students towards expert-like PS practices by encouraging them to follow a prescribed PS strategy that explicates the tacit PS processes of the expert including: 1) describing the problem situation in physics terms; 2) planning the construction of a solution; and 3) evaluation [1]. For PS practice to improve conceptual understanding, students should be encouraged to articulate their reasoning so they can self-explain how they applied domain concepts and principles to solve each problem [2]. Grading has a central role in shaping PS practices. Thus, within an instructional approach based on formative assessment, grading should reward explication of reasoning and the use of a prescribed PS strategy to help students learn from PS [5].

A central way to influence grading practices in a physics classroom is through graduate TAs, both because TAs are often responsible for grading students' work and because TAs are often required to participate in a professional development (PD) program. These PD programs should be based on research about the beliefs and practices of TAs.

As one piece of this research, we studied 43 graduate TAs entering their teaching career and a subgroup of 18 graduate TAs after a semester of teaching experience and a semester-long TA PD course. The PD course encouraged reflection on the various facets of teaching PS. In particular, we investigated the following research questions: What are TAs' grading practices and do they change after one semester of teaching experience and TA professional development? What are TAs goals and reasons for grading and how do those change? To what extent are the goals and reasons consistent with their grading practice?

## METHODOLOGY

Data collection took place at the beginning and end of the semester via a questionnaire designed to encourage introspection [6]. The 1st part of the questionnaire began with the general question: "What, in your view, is the purpose of grading students' work?" In the 2nd part, TAs were asked to make judgments about a set of student solutions to a physics problem (see Fig. 1). Here we focus on two of the five solutions (see Fig. 2). Clearly incorrect aspects of the solutions are indicated by boxed notes. For each solution, TAs were asked to complete a worksheet in which they listed characteristic features and explained how and why they weighed those features to obtain a specific score for both homework and quiz contexts. We focus here on the quiz context. We suggest that the reader examine the student solutions and think about how to grade them.

You are whirling a stone tied to the end of a string around in a vertical circle having a radius of 65 cm. You wish to whirl the stone fast enough so that when it is released at the point where the stone is moving directly upward it will rise to a maximum height of 23 m above the lowest point in the circle. In order to do this, what force will you have to exert on the string when the stone passes through its lowest point one-quarter turn before release. Assume that by the time that you have gotten the stone going and it makes its final turn around the circle, you are holding the end of the string at a fixed position. Assume also that air resistance can be neglected. The stone weighs 18 N.

**FIGURE 1**. Problem statement

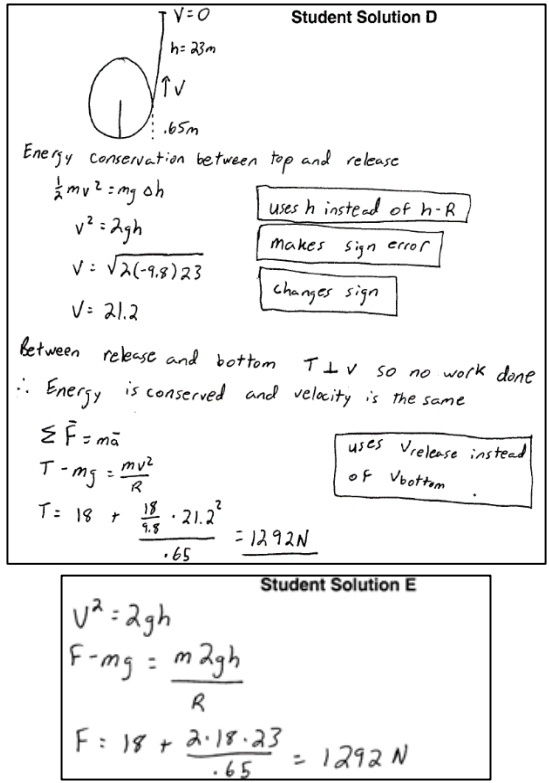

**FIGURE 2**. Student Solution E (SSE) and D (SSD)

The student solutions were selected to reflect expert and novice approaches to PS and to trigger instructional considerations related to encouraging (or not) expert-like PS approaches. For example, in comparing SSD to SSE, note that both include the feature of a correct answer. However, only SSD includes a diagram, articulates the principles used to find intermediate variables, and provides clear justification for the final result. In contrast, SSE is brief with no explication of reasoning. The elaborated reasoning in SSD reveals two canceling mistakes, involving misreading of the problem situation as well as misuse of energy conservation to imply circular motion with constant speed. SSE does not give any evidence of mistaken ideas, however, the student might be guided by a similar thought process as SSD. Thus, from a formative assessment point of view of encouraging prescribed PS strategy and explicit reasoning, SSD is somewhat better.

The data collection questionnaire also served as part of a TA PD course that was designed to encourage reflection on the various facets of teaching PS. TAs first completed the questionnaire individually and then discussed their grading practices and considerations in groups of three. These discussions often elicited conflicting viewpoints about the grading of the PS strategies and explication in the five solutions (in particular, SSD and SSE), during which the TAs attempted to resolve their conflicts. At the end of the class, the instructor coordinated a whole-group discussion in which the TAs shared their grading practices and considerations. Finally, the TAs completed the 2$^{nd}$ part of the survey individually, concluding the PD activities regarding grading.

## RESULTS - GRADING PRACTICE

*Result 1: The grading practices of TAs do not reward explication and the use of prescribed PS strategies. Moreover, these practices do not change significantly over the course of one semester of teaching experience and TA PD.*

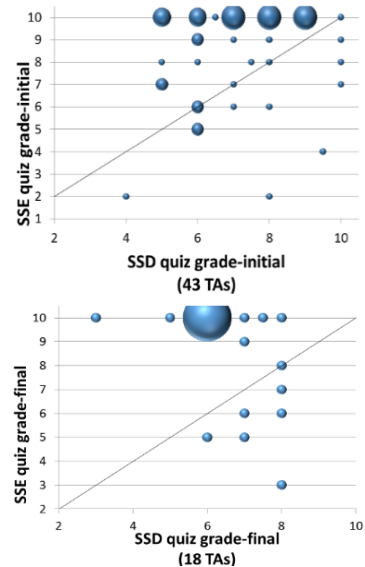

**FIGURE 4**. Distribution of TA grades, initial and final

At the beginning (initial) and end (final) of the semester, the majority of TAs grade SSE higher than SSD (see Fig. 4). The average grade of SSE and SSD of the subgroup of 18 TAs does not change significantly over the course of the semester (initial: <SSE>=7.7 and <SSD>=7.0, final: <SSE>=8.3 and <SSD>=6.6). Even after a semester of experience and

PD, TAs grade a solution which provides minimal reasoning (lacks effective PS strategies) while possibly obscuring physics mistakes higher than a solution which has detailed reasoning (and productive PS strategies) and includes canceling physics mistakes.

The lack of change in TAs' grading practices suggests that their practices are reflective of entrenched norms. To explore this possibility, we compared our results to a study by Henderson et al. [7] in which 30 faculty members were asked to grade SSE and SSD.

TABLE 1. Comparison of TAs and Faculty grading practices
Note: results for TAs after teaching experience and PD.

|  | SSE>SSD | SSE<SSD | <SSD> | <SSE> |
|---|---|---|---|---|
| Faculty | 43% | 40% | 7.7 | 8.0 |
| TAs | 61% | 33.3% | 6.6 | 8.3 |

Table 1 shows that faculty are more willing than TAs to reward SSD (the difference is statistically different, p-value: 0.002); faculty appear to appreciate reasoning and want to encourage it. However, the scores on SSE are remarkably similar (no statistically significant difference). This suggests that both faculty and TAs are not willing to penalize students who do not provide reasoning if their answer is correct. These core beliefs about grading are likely deeply held and resistant to change, even after PD and significant experience.

To understand why the TAs grade the way they do, we examined their goals for grading, the reasons they provide for the final grade they assign, and how their reasons relate to their grading practices.

## RESULTS - PURPOSE OF GRADING

*Result 2: TAs' stated purposes for grading students' solutions are to serve as a learning opportunity for the student and feedback for the instructor.* TAs' beliefs about the purpose of grading (from the open-ended question in the 1st part of the questionnaire; "What is the purpose for grading student solutions?") were coded using a bottom-up approach. TAs' general goals about the purposes of grading fell into four categories—to provide a learning opportunity for the student, to provide instructors with feedback on common difficulties of their students, to provide institutions with grades, and to motivate the students (e.g., to turn in their homework or to study harder).

At the beginning and end of the semester, almost all of the TAs state that grading serves as a learning opportunity for the student—to reflect on their mistakes and learn from them. Approximately half of the TAs state that it is for the benefit of instructor to understand student difficulties (See Table 2). However, our study suggests that TAs' stated goals are not aligned with their practice, as a majority of TAs grade SSE>SSD and transmit a message that explicit invocation and justification of principles are not important when solving physics problems. Additionally, solutions lacking reasoning do not provide evidence about students' thinking that can serve reflection and feedback.

TABLE 2. Responses to the purpose of grading before (initial) and after (final) teaching experience and PD.

| Purpose of Grading | | Initial (N=43) | Final (N=18) |
|---|---|---|---|
| For students | Learning opportunity | 93% | 100% |
| | Motivation | 21% | 33% |
| Feedback for instructor | | 58% | 39% |
| Grade for institution | | 16% | 39% |

## RESULTS - REASONS FOR GRADES

*Result 3: After teaching experience and PD, many of the TAs changed their reasons for assigning a specific grade due to a change in their perception of what counts as adequate evidence of students' thought processes.* TAs' reasons for the actual grade they assigned (TAs were asked to provide written reasons for the grades they assigned to each solution) were also coded using a bottom-up approach. We focus on SSE because few TAs mentioned reasons for the grade on SSD and they mostly focused on physics and math mistakes. The most common reasons for grading SSE were coded in four categories:

a) Adequate evidence – the TA can understand the student's thought process (e.g., "[SSE is] brief, but I can still understand what was done")
b) Inadequate evidence – the TA cannot understand the student's thought process (e.g., "he didn't prove that he understood the problem or accidentally [got it]")
c) Time/stress – there is limited time on quiz, so lenient grading is warranted (e.g., "in the quiz, in which time is limited, I will give full grade to this solution.")
d) Aesthetics – physics problems should be solved in a brief, condensed way (e.g., "The student had the right idea of how to approach the problem the simplest way. This approach is more preferable in quizzes because of its conciseness.")

TABLE 3. Reasons for the final grade on SSE before (initial) and after (final) teaching experience and PD.

| Reasons for Assigning a Grade | Initial (N=43) | | Final (N=18) | |
|---|---|---|---|---|
| | E>D | D≥E | E>D | D≥E |
| TAs giving reasons | 16 | 9 | 4 | 6 |
| Adequate evidence | 8 (19%) | 2 (5%) | 2 (11%) | 0 |
| Inadequate evidence | 2 (5%) | 7 (16%) | 0 | 6 (33%) |
| Time/stress | 5 (12%) | 0 | 2 (11%) | 0 |
| Aesthetics | 5 (12%) | 0 | 0 | 0 |

Table 3 shows that about 60% of the TAs provide reasons before (25/43) and after (10/18) teaching experience and PD. Initially, most TAs grade SSE>SSD whether they give reasons (16/25) or do not (12/18). In the final grading activity, the situation changes: of those who give reasons, most actually grade SSD≥SSE (6/10), while among those who do not provide reasons, most grade SSE>SSD (7/8). Initially, 7/43 (16%) penalize the lack of evidence in SSE, while finally 6/18 (33%) do so.

We examined more closely changes in the sub-group of 10 TAs who provided reasons. Four of these TAs scored SSE>SSD after teaching experience and PD. The average score given to SSE by these TAs increased from 6.5 initially to 9.8 finally. Half (2 TAs) changed from grading SSD>SSE to SSE>SSD. One of them initially revealed his conflict regarding evidence, stating, "It comes to mind that maybe the student cheated. But it is also possible that he/she did it him/herself." Afterward, this same TA resolved his conflict by considering time constraints, stating, "In quiz, maybe he/she didn't have time to write down everything." In the final stage, all four TAs mentioned adequate evidence and time constraints as reasons for the final grade on SSE. This suggests that teaching experience can influence TAs to require less evidence due to time constraints in the quiz context.

However, regarding the SSD≥SSE group (6 TAs at the end of the semester), the average score given to SSE by these six TAs dropped from 6.8 initially to 5.3 finally. Half of the group (3 TAs) initially scored SSE≥SSD. Initially, only two of the six TAs mentioned the reason of inadequate evidence, while others said that SSE contained the correct answer. One of the six TAs initially mentioned that SSE was aesthetically correct, stating: "I love this…solution. No extra words." After experience and PD, this same TA said, "There is no description…it's too hard to follow." The three TAs who initially scored SSE≥SSD switched from mentioning reasons of correctness and aesthetics to inadequate evidence and the number of TAs mentioning inadequate evidence increased to six. This suggests that teaching experience and PD can influence TAs to require more evidence in the quiz context.

In general, there is no definite direction in the shift of preferring SSD to SSE (or vice versa). However, there is a trend in that some TAs change their reasons when assigning a specific grade, in particular with regard to what they perceive as adequate evidence in students' solutions.

## DISCUSSION AND SUMMARY

This study found that after one semester of teaching experience and a semester-long PD intervention which was designed to encourage reflection on the various facets of teaching PS, TAs:

- *Maintained a grading practice that does not reward explication and the use of prescribed PS strategies.*
- *Maintained general goals for grading – to provide a learning opportunity for the student as well as to provide instructors with feedback on common difficulties of their students.* However, for most of the TAs, these goals are not supported by their actual grading practices.
- *Changed their reasons when assigning a specific grade, in particular with regard to what they perceive as adequate evidence in students' solutions.*

The main limitation of this study is due to the relatively small number of participants (18 at the end of the semester). Even in light of this limitation, the results of this study can inform PD in preparing TAs for their grading responsibilities. It cannot be expected that TAs' grading practices would encourage use of prescribed PS strategies and explication as the TAs gain teaching experience. In fact, the TAs' grading practices do not seem to change much, if at all, after a semester of teaching experience and PD, as most TAs continue to grade SSE>SSD. Since most TAs already view grading as a learning opportunity for both the student and the instructor, PD should focus on building on this productive belief about the purpose for grading. Moreover, PD can build on the instability in TAs perception of adequate evidence in students' solutions, directing TAs to better match their goals and practice via a more critical examination of their perception of evidence and how it conflicts with other considerations, such as time constraints or aesthetics. For example, PD could elicit conflicts between beliefs about the purpose of grading and actual grading practices and stir discussion about the tension between evidence and time constraints in assigning quiz grades and the messages sent to students.

## REFERENCES

1. F. Reif, *Applying Cognitive Science to Education: Thinking and Learning in Scientific and Other Complex Domains* (MIT Press, Cambridge, MA, 2008).
2. M. Chi, in *Advances in Instructional Psychology*, edited by R. Glaser (Lawrence Erlbaum, Mahwah, NJ, 2000).
3. D. Maloney, in *Handbook of Research in Science Teaching and Learning*, edited by D. Gabel (MacMillan, 1993).
4. E. Yerushalmi and B. Eylon, in *Encyclopedia of Science Education*, edited by R. Gunstone (Springer, 2014).
5. P. Black and D. Wiliam. Phi Delta Kappan **80**(2), 139 (1998).
6. E. Yerushalmi, et al. PRST PER **3,** 020109 (2007).
7. C. Henderson et al., Am. J. Phys. **72**, 164 (2004).